\documentclass[12pt]{article}
\usepackage{epsfig}
\include{fig1.eps}
\include{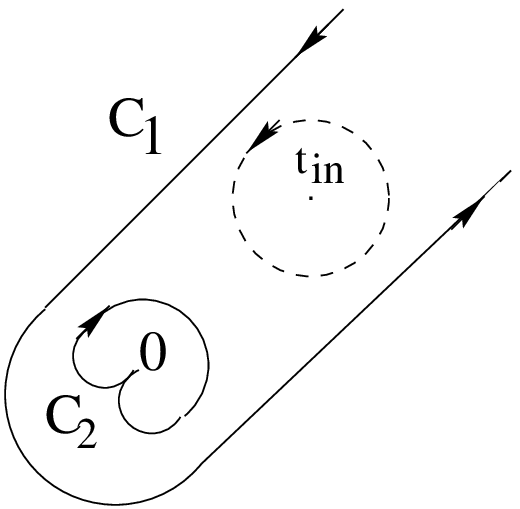}
\renewcommand{\theequation}{\arabic{section}.\arabic{equation}}
\newcommand{\cleqn}{\setcounter{equation}{0}}
\newcommand{\sectionnew}[1]{\section{#1}\cleqn}
\def\be{\begin{equation}}
\def\ee{\end{equation}}
\def\bea{\begin{eqnarray}}
\def\eea{\end{eqnarray}}
\def\ra{\rightarrow}
\def\lra{\leftrightarrow}
\def\al{\alpha}
\def\ga{\gamma}

\def\grad{\rm grad}

\def\th{\theta}
\begin{document}
\begin{center}
{\large\bf The Aharonov-Bohm Problem Revisited}\\
\vspace{1cm}
Yoichiro Nambu \\
\vspace{.5cm}
The Enrico Fermi Institute\\
University of Chicago \\
\end{center}
\vspace{1.5cm}
\begin{abstract}
The properties of a nonrelativistic charged particle in two-dimensions in the 
presence of an arbitrary number of nonquantized magnetic fluxes are 
investigated in free space as well as in a uniform magnetic field. The fluxes 
are represented mathematically as branch points in one of the complex@
coordinates. To construct many-flux solutions without a magnetic field, 
however, the fluxes must be treated as dynamical objects dual to the charges. 
A medium made up of fluxes acts like an anti-magnetic field and tends to expel 
the charges. 
\end{abstract}
 
\bigskip
 
\sectionnew{\large Introduction}
 
In a classic paper, Aharonov and Bohm \cite{AB} have pointed out
that a locally trivial vector potential of a gauge field can lead
to observable effects in a nontrivial topology through the phase of
the wave function of a charged particle. This has spawned many
important theoretical concepts in later developments in gauge
theories, like instanton, monopole, and confinement by
monopole condensation. On the experimental side, confirmation of
the original Aharonov Bohm (A-B) effect due to a nonquantized
magnetic flux tube has been somewhat controversial because of
the difficulty of setting up ideal conditions, but there have
been several experiments, notably those by Tonomura {\it et al}
\cite{Tonomura} showing the shifts in diffraction fringes due
to magnetic fluxes, as expected by the A-B effect. (See \cite{Olariu} 
for a general review of the problem.)  The present work was motivated 
by the theoretical question: What would be the properties of a medium 
filled with many nonquantized magnetic fluxes (vortices)? Assume, for 
simplicity, that the fluxes are all parallel so that problem can be 
reduced to that of an electrically charged particle in a 2-plane pierced 
by pointlike magnetic fluxes. Then, for example, what would be the 
behavior of a charged particle in a lattice of such fluxes? What 
would be its energy spectrum? Would the fluxes cause more drastic
effects than a phase shift in the wave function? Although the
semiclassical arguments suffice to discuss and analyze most
situations realistically, it requires a more rigorous treatment
of the Schroedinger equation to answer these questions. This
turns out to be a highly nontrivial problem. To our knowledge, the 
solution of the Schroedinger equation in the presence of two or more 
fluxes has not been explicitly constructed. (A qualitative discussion 
of special cases was made by Peshkin et al. \cite{Peshkin}. Some other 
theoretical problems related to the present one have been discussed by 
Aharonov and Casher \cite{AharonovCasher}, Dubrovin and 
Novikov\cite{Dubrovin}, Jackiw\cite{Jackiw}, and Lewis\cite{Lewis}.
 
We develop here a method of solving the many-flux problem by
adapting the work of Sommerfeld\cite{Sommerfeld}\cite{Pauli}. He 
solved the problem of diffraction of light by a semi-infinite wall in 
two dimensions by regarding the wall as a branch cut in a two-sheeted
Riemann surface. The incoming plane wave in the physical plane
dives into the unphysical second sheet as it hits the cut, while
the reflected wave emerges from the second sheet through the
cut. The solutions were constructed as a contour integral  of a
kernel in a complex angular variable over the two sheets.
 
The A-B problem can be posed in a similar fashion. The pure
gauge vector potential of a flux is singular at the flux site.
In the complex variables $z=x+iy$ it behaves like
($\al/2)(i/z,-i/\bar{z})$, where $2\pi\al$ is the flux strength. After
removing the potential by a gauge transformation, we get a wave
function that satisfies a free Schroedinger equation, but it is
singular in the sense that the function acquires a phase 
$\eta = \exp(2i\pi\al)$ after encircling the flux once, so the wave
function must vanish at the flux if $\al$ is noninteger.
Regarded as a function in the complex $2$-plane, the problem then
reduces to that of solving the free Schroedinger equation on a
Riemann surface where the flux is a branch point, and the
solution must satisfy the boundary condition that it gets a phase
change $\eta$  when going around it once, and vanishes at the
branch point.
 
The paper is organized as follows. We first establish the basic
formalism of complex integral representation, apply it to
rederive the known results for the case of a single flux with or
without the presence of a uniform magnetic field.\cite{AB}\cite{Lewis}. 
Then the problem of many fluxes is addressed by two methods which 
yield different types of solutions. One of them is to use multiple 
integral representations, and is applicable only if a magnetic field 
is present. The other is to treat the fluxes as dynamical objects 
having properties dual to the charges. An important general property 
that emerges is that the nontrivial phase information around all fluxes
must be represented as branch points in either the $z$ or the $\bar{z}$ 
coordinate, but not a mixture of them. It means that, since the wave 
function must vanish at the flux sites, the (fractional) angular 
momentum around each flux, as opposed to the intrinsic flux strength, 
has the same sign, either positive or negative. Thus for example, 
if the two flux strengths are $\al_1> 0$ and $\al_2 < 0$, the solution 
is constructed as having positive angular momenta $\al_1+l_1>0$ and 
$l_2+\al_2>0,l_i=$ integer, in the $z$ variable (and similarly negative 
angular momenta in $\bar{z}$). Otherwise the solution would have to be
represented by a function in $z$ in a region around one flux, and by
a function in $\bar{z}$ in another, but it is not possible to match
the values of the wave functions and their derivatives at the
interface.
\footnote{To see this, consider at zero energy two fluxes of
opposite sign located at $\pm ib$. Since any analytic function
of $z$ or of $\bar{z}$ satisfies the Shroedinger equation, assume:
$\psi = f_1 = (z-ib)^{\al}g(z)$ in the upper half plane
($y=\Im z > 0$), where $g(z)$ is regular, and %
$\psi = f_2 = (\bar{z}-ib)^{\al}g(\bar{z})$ in the lower half
plane 
imposed, and analytically each can be extended to the whole plane. 
Obviously $f_1 = f_2$ on the real line, but their normal derivatives 
are opposite and $\neq 0$. (If the derivatives were zero, the functions
would vanish identically.) The same conclusion is reached if the
solution in each half plane is itself a sum of analytic and 
antianalytic parts.}
 
The fact that angular momenta around all the fluxes must have the same sign
is not intuitively clear, but may be inferred from the zero energy solutions, 
where the nontrivial information about all the fluxes must be encoded in
terms of either analytic or antianlytic functions. It is also understandable 
from the following consideration. Suppose $\al_1=-\al_2$, and let them approach
each other and collapse to nothing. The wave function originally vanished at 
both sites and was finite elesehwere. So it will vanish at the point of 
collapse, hence it will be in a $p$ or higher angular momentum state.

One of the consequences of the above property is that a medium
made up of fluxes tends to expel the charge as if it was placed
in an anti-magnetic field. This behavior has a simple explanation. 
By construction the angular momentum around each flux is of the same 
sign, say $\al'$, irrespective of the sign of its intrinsic flux strength. The 
total angular momentum around a circle at a radius $R$ lattice units will 
be $\sim\al' \pi R^2$, hence the radial momentum $p(R)$ is 
$\pm (E - (\al'\pi R)^2)^{1/2}$. For large $R$, then, the wave function will go
like $\exp(\int pdR)\sim\exp(\pm \al'\pi R^2/2)$. The minus sign is excluded 
by the boundary condition at the flux sites
 
\bigskip
 
\sectionnew{\large Basics}
 
Let a flux of strength $\al$ be located at the origin. For an `electron' of 
charge -e, the Aharonov-Bohm gauge potential $A$ and the covariant 
derivative $D$ in the standard circular gauge are given by
\bea
(A_x,A_y)&=&(\al/2)(y/r^{2}, -x/r^2)\nonumber\\
&=&(\al /2)(\partial_{y}, -\partial_{x})\ln(r), \nonumber \\
\nabla \times A&=&2\pi \al \delta^2(x), \nonumber \\
D \equiv (\nabla - iA)&=&(\partial/\partial{x} +iy/(2r^2),
\partial/\partial{y} -ix/(2r^2))
\label{2.1}
\eea
For notational convenience, we are taking $eh=1$ so that
integer $\al$ corresponds to a quantized flux.
In terms of the complex variables
$z = x + iy$ and $\bar{z} = x - iy$, they become
\bea
&&x = (z + \bar{z})/2, \, y = (z - \bar{z})/2i, \nonumber \\
&&\partial/\partial{z} =
(1/2)(\partial/\partial{x} - i
\partial/\partial{y}), \nonumber \\
&&\partial/\partial{\bar{z}}= (1/2)(\partial/\partial{x} + i
\partial/\partial{y}), \nonumber \\ &&(\partial^2/\partial{x^2}
+ \partial^2/\partial{y^2}) =
4(\partial/\partial{z})(\partial/\partial{\bar{z}} ),\nonumber\\
&&A_z = i\al/2z, \, \, A_{\bar{z}} = -i\al /2\bar{z} \nonumber
\\
&&D_z=\partial_z+\al/(2z),\,\, D_{\bar{z}}=\partial_{\bar{z}}-\al/(2\bar{z})
\label{2.2}
\eea
The basic Schroedinger equation reads, in time dependent (energy 
Eigenvalue) forms, after redefining $mE/2 = k^2/4 \ra E$
\bea
(\{D_z,D_{\bar{z}}\}/2 +\partial_{\tau})\Psi&=&0 \nonumber\\
((\{D_z,D_{\bar{z}}\}/2 + E)\Psi&=&0)
\label{2.3}
\eea
with the condition that $\psi$ is one-valued and finite. (We
will not necessarily require finiteness of derivatives, though.)
Here a Euclidean time is used for later convenience, $z$ and $\bar{z}$ are
regarded as independent variables spanning a complex 2-dimensional plane. 
The physical space is its subspace where z and $\bar{z}$ are complex 
conjugates of each other (real $x$-$y$ plane). Hereafter we will refer 
to this situation as "on shell". Since Eq.(2.3) has independent 
first derivatives in $z$ and $\bar{z}$, the actual physical space also 
includes its tangential neighborhoods. After solutions have been written down 
in $z$ and ${z}$, however, we can stay on shell by reverting to $x$ and $y$.
The ordering of $D_z$ and $D_{\bar{z}}$ may be ignored since their potential 
noncommutativity at their singularities will not arise. From Eq.(2.2) we see 
that Eq.(2.3) is invariant under the operation
\be
z \lra \bar{z}, \hskip10mm \al  \ra  -\al
\label{2.4}
\ee
It corresponds to a reflection $y\ra -y$.  The interchange of $z$ 
and $\bar{z}$ is also effected by complex conjugation of the wave function 
which corresponds to time reversal.
 
We next eliminate from Eq.(2.3) the gauge potential by a
singular gauge transformation $G$:
\be
\Psi  =  G\psi, \,\, G = (\bar{z}/z)^{\al/2},
\label{2.5}
\ee
so that $\psi$ now satisfies a free Schroedinger equation.
\bea
(\partial_{z}\partial_{\bar{z}} +\partial_t)\psi&=&0 \nonumber\\
((\partial_{z}\partial_{\bar{z}} + E)\psi&=&0)
\label{2.6}
\eea
($A$ is pure gauge if its $z (\bar{z})$ component is a function of 
$z(\bar{z})$ only. See Appendix 1.) Since the original $\psi$ is 
one-valued, $\psi$ must be singular in such a way as to cancel the 
singularity in the gauge function in Eq.(2.5). This means that 
$\psi$ must be defined as a free wave function on a Riemann surface 
with a branch cut running from $0$ to $+\infty$, and on shell there 
is a phase change as we go around the origin once:
\bea
&&\psi(\theta=2\pi)=\eta\psi(\theta=0),,\,\ \eta=\exp(2i\pi\al),\nonumber\\
&&\psi(z = \bar{z} = 0) = 0
\label{2.7}
\eea
and subject to conditions of finiteness at infinity on shell.
In the following we will mainly be concerned with the eigenvalue equation. 
An elementary solution of Eq.(2.6) is
\be
\exp(zt-E\bar{z}/t)
\label{2.8}
\ee
where $t$ and $\bar{t}$ are complex momenta $t$.  
The general solution of Eq.(2.6) that satisfies the boundary conditions 
may be built up as a superposition  of elementary solutions
\be
\psi = \int\exp(zt- \bar{z}/t)f(t)dt
\label{2.9}
\ee
with an appropriate choice of the function $f(t)$ and the integration 
path in such a way as to satisfy Eqs.(2.7) on shell. We recognize the familiar 
integral representation of Bessel functions. 
 
Now assume for the time being  $0 < \al < 1$ , and let
\be
f(t) = t ^{-\al - n -1},  \hspace{10mm}  n = 0,1,2,\ldots
\label{2.10}
\ee
and choose the contour $C$ of integration to be
\be
C:   (U(\Re(zt) < 0)\infty, (0+)),
\label{2.11}
\ee
which goes around $0$ starting from, and ending at, infinity in the direction 
$U$ such that the integral is convergent. This means that, as the phase of 
$z$ rotates by $2\pi$, the contour integration will have to make a 
counter-rotation by $-2\pi$, so the factor $f(t)$ will yield a phase factor 
$e^{2i\pi\al}$ as is required by Eq.(2.7). Furthermore, $\psi'$ stays finite 
on shell when $z$ and $\bar{z}$ go to infinity, and also at zero since
\be
\int t^{-\al -1- n}dt = 0,     \, \, n + \al > 0
\label{2.12}
\ee
These properties can be made more explicit by a change of integration 
variable, $t \ra E/z$
\be
\psi=z^{n+\al}\int_{C}{\exp(Et-z\bar{z}/t)t^{-\al-n-1}dt}
\label{2.13}
\ee
up to a constant factor, and where the contour $C$ is now
$(-\infty, (0+))$. So Eq.(2.13) gives a desired set of solutions
for any $\al$ and integer $n=0,1,2,\ldots$ that satisfy $n+\al > 0$. 
In the case $n + \al< 0$, we can choose a new contour around $0$:
\be
C': (0 \times U(\Re{1/t} > 0, (0+)),
\label{2.14}
\ee
i. e.\, one that starts from the origin in the direction $U$, and comes 
back after encircling it clockwise. This is equivalent to the conjugate 
form ($\bar{z}$-type) of $\psi$ according to Eq.(2.4), which can be 
converted to the original contour by the substitution in Eq.(2.10): 
$t \ra E/t$,  resulting in a new $f(t)'= t^{\al + n-1}$. Eq.(2.13) then 
becomes, up to a constant factor,
\be
\psi=\bar{z}^{-\al-n}\int_C{\exp(-Et+z\bar{z}/t)
t^{\al+n-}dt},\,n+\al<0
\label{2.15}
\ee
In terms of Bessel functions, the $z$- and $\bar{z}$-type
solutions lead to solutions in the original gauge:
\bea
&n+\al>0, z{\rm -type}:&\Psi =
(z/\bar{z})^{n/2}J_{\al+n} (2(Ez\bar{z})^{1/2}),\nonumber\\
&n+\al < 0,\bar{z}{\rm -type}:&\Psi = (z/\bar{z})^{n/2}J_{|\al+n|}
(2(Ez\bar{z})^{1/2})
\label{2.16}
\eea
Thus the physical states are labeled by $|\al + n |$ and an integer angular
momentum $n = 0,\pm 1,\pm 2, \cdots$. The angular momentum operator $L$ is 
given by $z\partial_z - \bar{z}\partial_{\bar{z}}$. Eqs.(2.13) and (2.15) show 
that the nontrivial phase information in $\psi$ is carried respectively by $z$ 
and $\bar{z}$. Applying the operators $\partial_z$ and $\partial_{\bar{z}}$ on 
a $\psi$ serve to change $n$ by $\mp 1$. This may be repeated any number of 
times as long as the wave funtion remains finite, thereby generating a complete
set of states.

Some remarks are in order here about the problems associated with 
non-single valued wave functions and the limitations in taking derivatives. The
first problem poses a difficulty in defining a complete set of states. 
Although scalar products are independent of the choice of (unitary) gauges, it 
does not make sense to expand a function with a branch point at $z=a$ in terms 
of functions with a branch point at $z=b$. Similarly, the meaning of going to 
momentum space by Fourier transforms is not clear since it will depend on where
a cut is made, and is diffucult to evaluate. We should, therefore, go back to 
the original non-singular gauges to build a Hilbert space of physical states. 
We may, however, work with non-single valued functions in enumerating the 
states since there is a 1-1 correspondence between the two sets. 

The second problem, on the other hand, causes a problem in defining higher 
powers of mementum as observables, e.g., the finite translation operators in 
power series. For convenience, however, we may pretend to ignore these 
problems, and use such formal expressions as $(i\bar{k})^{\al}$ to mean 
$\partial_z^{\al}=t^{\al}$ when applied to Eq.(2.8) in momentum space.

\bigskip
 
\sectionnew{\large Scattering}
 
Since the solutions Eqs.(2.16) are Bessel functions of non integer order,
we expect that the scattering amplitude of an electron by the flux can
be obtained by a superposition of all solutions labeled by $n$, which
would form a complete set. However, since they all vanish at the origin,
they form a complete set only in the punctured plane $R^2 - 0$, hence
it would not be possible to construct a plane wave in $R^2$ with this
set of functions. What is missing is an equivalent of the $s$ wave
component that remains finite at the origin, a situation analogous to
the case of hard sphere scattering. But it is also different
because of the long range nature of the gauge potential as in
the case of the Coulomb scattering.
 
A proper scattering amplitude can be constructed in the following way.
Consider the integral.
\bea
&&\psi = \int_{C_1 + C_2}{\exp(Ezt-\bar{z}/t)f(t)dt}, \nonumber\\
&&f(t)= (t/t_{in})^{\al}/(t-t_{in}), \hspace{1cm} t_{in}= iE^{1/2}
\label{3.1}
\eea
$t_{\rm in}$ is the momentum of the incoming plane wave moving
in the $x$ direction:
\be
\psi_{in}= \exp(iE^{1/2}(z +\bar{z}))= \exp(2iE^{1/2}x)
\label{3.2}
\ee
The contours $C_1$ and $C_2$ are defined as follows. [Fig.1]
Denote by $R_0$ the reference Riemann sheet, and by $R_n$ the sheet
reached by going counter-clockwise around the origin $n$ times.
For a given position $z = r\exp(i\theta) = rU, C_1$ starts from
$\infty\times U^{1}$, goes counterclockwise around the origin
and $t_{in}$, to end up at $\infty\times U^{1}$ on $S_1$. $C_2$
starts from $0 \times U^{-1}$, goes clockwise around the origin,
but avoiding $t_{in}$, to end up at $0\times U^{1}$.\\

\bigskip 

\centerline{\epsfbox{fig1a.eps}}

\centerline{{Fig.1 Contours for scattering}}

\bigskip

Some properties of Eq.(3.1) are obvious:\\
1) The integral converges for $z = \bar{z}= 0$, giving $\psi =
0$.\\
2) When $\al$ is an integer, the two contours can be detached
from their limits to form a closed circle around $t_{in}$, which
then picks up the incoming wave $\psi_{in}$ only, so there is no
scattering.\\
That the formula gives the correct scattering amplitude can be
shown by computing its asymptotic form by the saddle point
method. First take the general formula Eq.(2.9). Its saddle
points are determined by the extrema $t_0$ of $S(t)$, and
performing a Gaussian integration around them in the direction
of the steepest descent. In the case of scattering,
\bea
&&S' = z + E\bar{z}/t_0^2 + \al/t_0 - 1/(t_0-t_{in}) =
0,\nonumber\\
&&S" = -2E\bar{z}/t_0^3 - \al/t_0^2 + 1/(t_0-t_{in})^2
\label{3.3}
\eea
For large $E^{1/2}|z| = E^{1/2}r >>1$, there are three
saddle points on a circle of radius $E^{1/2}$: one near the pole
$t_{\rm in}$ representing the plane wave, and a conjugate pair
$t_{+}, t_{-}$ representing radially outgoing and incoming waves
respectively. ($t_{in}$ and $t_{\pm}$ do not coincide except
when $\theta = 0$ or $\pi$.) But the contours. $C_1$ and $C_2$
are such that the incoming component is canceled, leaving only
the outgoing one. For the radial components,
\bea
&&t_{\pm} \sim E^{1/2}\exp(\pm i\theta), \nonumber \\
&&e^S \sim -iE^{-1/2}\exp(\pm 2iE^{1/2}R-i\al\theta\pm
i\pi/4)/(\pm e^{-i\theta}-1)
\label{3.4}
\eea
The Gaussian integration around $t_{\pm}$ yields
\bea
&&\psi_{+-} \sim -\exp(\pm 2iE^{1/2}R-i(1/2+\al)\theta \pm
i\pi/4))/R^{1/2} \nonumber\\
&&/\sin(\theta/2)  \hskip10mm {\rm for}\, \psi_{+}, \nonumber\\
&&/\cos(\theta/2)) \hskip10mm {\rm for}\, \psi_{-}
\label{3.5}
\eea
These should also be multiplied respectively by an additional
factor depending on the contour: $-\eta (\eta^{-1})$ and $-1(1)$
for $C_1 (C_2)$. So when the two contributions are added, and after
reverting to the original gauge, the scattering amplitude becomes
\be
\psi_{scat} \sim \sin(\pi\al)\exp(2iE^{1/2}R-i(1/2+\al)\theta
-i\pi/4)))/(R^{1/2}\sin(\theta/2))
\label{3.6}
\ee
which reproduces the result given by \cite{AB}. The formula is
valid except for a near forward region where the width of the
Gaussian integration $\sim R^{1/2}E^{1/4}$ becomes larger than
$|t_{in}-t{+}|$. (see \cite\cite{Ruijsenaars}\cite{Jackiw}{Sakoda}).

\bigskip
 
\sectionnew{\large Presence of a magnetic field}
 
\subsection{Basic formulas}
 
In this section we allow for the presence of a real magnetic
field in addition to the fluxes. The vector potential for a
constant magnetic field $\ga$ is 
\be
A_z = \ga\bar{z}/2, \,\, A_{\bar{z}}= -\ga/z/2
\label{4.1}
\ee
which is to to be added to the Aharonov-Bohm potential. After
gauging away the latter, we have
\be
H\Psi=E\Psi, H=-\{(\partial_z-\ga\bar{z}/2),(\partial_{\bar{z}}+\ga
z/2)\}/2,
\label{4.2}
o\ee
Assuming $\ga > 0$, make a further transformation
\bea
\Psi&=&\exp(-\ga z\bar{z}/2)\psi,\,\,H'\psi= E\psi,\nonumber\\
H&=& (\partial_{z}\ga\bar{z})\partial_{\bar{z}}+\ga/2
\label{4.3}
\eea
Since $\partial_z$ commutes with $H$, an elementary solution is
then
\be
\psi =  \exp(zt)(t - \ga\bar{z})^{E/\ga - 1/2},
\label{4.4}
\ee
where $t$ is a complex constant. If there are no fluxes, $\psi$
must be single-valued and finite everywhere, i.e., the second
factor must be a polynomial, which leads to the familiar Landau
spectrum:
\be
E/\ga -1/2 = n = 1, 2 \cdots
\label{4.5}
\ee
so that
\be
\psi_n =  \exp(-\ga z\bar{z}/2 + tz)(t - \ga\bar{z})^n
\label{4.6}
\ee
For $n \ge 0$, $\psi$ vanishes at  $\ga\bar{z} = t$, and goes to
zero at large $|z|$.  The states are labeled by the level number
$n$, and a continuous complex parameter $t$ which is the analog
of the continuous real parameter in the Landau gauge. (The true
energy is $E/4m$ in our convention.) Here $t/\ga$ corresponds to
the center of the classical Larmor orbit with angular momentum
$-n \leq 0$. Note that $(t-\ga\bar{z})$ is a raising operator for
the level number $n$, hence $<t> = <\ga\bar{z}>$. We will
conveniently refer to this factor as raising factor.
 
The space of all $n$'s and all complex $t$'s is overcomplete on
the physical shell, unlike in the Landau gauge. (In the circular
gauge, the radius $|t|$ and $n$ label the states \cite{Lewis}.) But by 
expanding Eq.(4.4) in powers of $t$, or by the Fourier integral
\be
\psi_{n,l} = \oint{\psi_n t^{-l-1}dt},  \, \, l \geq 0
\label{4.7}
\ee
we get a complete set centered at zero and labeled by two
integers $n$ and $l$, giving a total angular momentum $j = l-n$.
Each solution contains a factor $z^l$ times a polynomial of
order $n$ in $z$ and $\bar{z}$. Eq.(4.7) allows an intuitive
interpretation that these states are an epicycle-like
superposition of Larmor orbits, with `orbital' angular momentum
$l$, along another circle of a certain radius (see below). When
$\ga < 0$, clearly we may take the conjugate solutions to
the above $z \lra \bar{z}, \ga \ra -\ga, j \ra -j$, and the
energy levels are given in general by $E/|\ga| = n + 1/2$. It is
also instructive to examine the limit $\ga \ra 0$, where we
should recover the original free field results. Since the energy
spectrum is proportional to $\ga$, however, in general it will
be driven to zero in this limit, as may be seen from the
elementary solution Eq.(4.6) which does not have a limit
$E=n/\ga$ except for $n=0, E \ra0$. But supplying a factor
$t^{-n}$, i.e.\, dividing the raising factor by $t$, we get
\be
\lim \hspace{1em}\exp(-\ga z\bar{z}/2 +zt)(1-\ga\bar{z}/t)^{E/\ga-1/2}
=\exp(zt-\bar{z}E/t)
\label{4.8}
\ee
which reproduces the free field exponent.
 
\bigskip
 
\subsection{A flux in a magnetic field}
 
There are two ways to introduce a flux of strength $\al$. One is
to change $n$ to $n+\al$ in Eq.(4.6) (provided that it is
$\geq 0$), and to let it be the gauge transformed $\psi$. It
represents a flux located at $t/\ga$, and the energy is shifted:
\bea
&&\psi_{n+\al,t} = \exp(zt)(t - \ga\bar{z})^{n+\al}, \nonumber\\
&&E/\ga - 1/2 = n +\al \geq 0
\label{4.9}
\eea
By expanding Eq.(4.9) in powers of $t$, or taking the moments
\be \psi_{n,l} = \oint{\psi_{n+\al,t}t^{-l-1}dt}, \, 0 \leq l
\leq l_0 = [n+\al]
\label{4.10}
\ee
we get $l_0$ states with the same shifted energy. The
restriction $l \leq l_0$ arises from  the condition $\psi(0)= 0$.
This is a $\bar{z}$ type  representation.
\footnote{It can be turned into a $z$-type representation if we change
the sign of $\ga$, and let the path be $C=(\infty,(0+))$. The
energy $\sim E/\ga$ remains positive. See section 6.3}
Both Eq.(4.9) and Eq.(4.10) are now nonpolynomial functions in $\bar{z}$.
 
For $l \ge l_{0}$, the proper formula is given by
\bea
&&\psi_{n,l} = \int_{C}{\psi_{n,t}t^{l - \al -1}dt}, \nonumber\\
&&E/\ga  - 1/2 =n,\hspace{1cm} C = (\infty,(0+, \ga\bar{z}+))
\label{4.11}
\eea
which is a $z$-type representation. The energy does not get
shifted for them. For $n = 0$, in particular, $\psi$ is the same
as the zero energy limit of the field-free case, i. e., a
product of $z_i$'s, apart from an overall Gaussian factor. The
reason why a nonpolynomial raising factor in the integrand is
not allowed is that, at large distances, $\psi$ becomes $\sim
\exp(+\ga z\bar{z}/2)$, as may be seen by finding the saddle
points. It is not possible to avoid the exponential blowup by
changing the sign of $\ga$ or energy. The problem is avoided
only if the factor is a polynomial. (According to the previous
subsection, it can also be seen that only unshifted levels have zero
field limits.) Obviously these same results will hold for $\ga< 0$,
for which the conjugate representations are used.
 
Summarizing, a number $0 \leq l_0 =[n+\al ]$ of states at level $n$
have their energy (more precisely $E/|\ga| -1/2$) shifted to
$n + \al \geq 0$, and the rest is uninfluenced by the flux. Consider
in particular the ground states $n=0$ with $|al| < 1$. If
$\al >0, \ga >0$ (or if $\al < 0, \ga < 0 $), i. e., the magnetic
field and the flux are parallel, then the energy of one state is
shifted upwards by $\al$. If, on the other hand, the field and the
flux are antiparallel, none of the ground states can change energy;
the energy can only go up, but not down below the zero point
minimum. 

The following semiclassical argument explains the
difference between shifted and unshifted levels (see also
\cite{Lewis}). The state $\psi_{n,t}$ around $t$ occupies a disk
of radius $R \sim (n/\ga)^{1/2}$. The orbital radius of the
`epicyle' state $\psi_{n,l}$ is
$r \sim l/E^{1/2}\sim l/(\ga n)^{1/2}$. So the latter state will
cover the flux at the origin only if $R < r$. As for the asymmetry
in the ground state, write the Hamiltonian as
\be
H = p^{2} - L (\al/r^{2}
+ \ga) + (\al/r + \ga r)^{2}
\label{4.12}
\ee
For $L = 0$, the minimum of the last term is $\al \ga$ if
$\al \ga \ge0$, and $0$ if $\al\ga < 0$.
 
\bigskip
 
\sectionnew{\large The multi-flux problem}
 
\subsection{Multiple integral representation}
 
At zero energy, the general solution of the Schroedinger equation
is either analytic or antianalytic. The one vertex solution
reduces to $\psi \sim z^{\al}$ or $\bar{z}^{\al}$. More than one
vertices can be easily accommodated by taking products of $n$
such $z$-type or $\bar{z}$type factors, but we cannot form
general mixed products. When $E>0$, the solutions will involve
both $z$ and $\bar{z}$, but by continuity this property will
persist. In the following we assume nonzero energy. Our goal is to
find a solution which has proper monodromy properties around each
flux:$M_i\psi= \eta_i\psi$, where $M_i$ denotes the monodromy
operation around flux $i$, giving a phase $\eta_i=\exp({i\th_i})$. Since,
however, the radial second order differential equation obtained after
removing $G$ has singularities only at $0$ and $\infty$, it is not clear 
how to introduce more flux singularities.

So let us first try a peturbative solution in energy for two fluxes. Denote 
their location and strength by $(b_1, \al_1), (b_2,\al_2)$, and write 
$z - b_1 =z_1, z - b_2 =  z_2$ for short:
$\psi=\sum \psi^{(n)}(z,\bar{z}), \psi^{(0)}=(z_1)^{\al_1}(z_2)^{\al_2}$.
Then
\be
E\psi^{(n+1)}=-\partial_z\partial_{\bar{z}}\psi^{(n)}
\label{5.1}
\ee
$\psi^{(1)}$ can be obtained by integrating $\psi^{(0)}$ over 
$z$ from $b_1\,\, (b_2)$, and over $\bar{z}$ from $\bar{b_1}\,\, (\bar{b}_1)$. 
This $\psi^{(1)}$ vanishes at both sites, but it does not have a proper 
monodromy at the second (first) site. We can make $\psi{(1)}$ behave correctly 
at both sites by taking $\psi^{(1)}=(z_1)^{\al +n_1}(z_2)^{\al_2 +n_2}\bar{z}$ 
with integer $n_i\geq 1$. $\psi^{(0)}$ is then obtained by differentiating 
$\psi^{(1)}$. But to generate a series to the $N$th order, we have to start 
from an $N$th order term of sufficiently high power $n_i\geq N$, and we lose 
copntrol of the perturbation expansion. 

Next we look for a nonperturtive solution in terms of a double integral 
representation for two fluxes.  The gauge potential and gauge function are 
respectively a sum and a product of two components:
\bea
A_z&=&\al_1/z_1+\al_2/z_2,\,\, A_{\bar{z}}=\bar{A}_z, \nonumber \\
G(z)&=&(\bar{z}_1/z_1)^{\al_1/2}(\bar{z}_2/z_2)^{\al_2/2}
\label{5.2}
\eea
Consider a solution of the Schroedinger equation of the form
\bea
&&e^S = \exp[z_1 t_1 + z_2 t_2
- E^{1/2}(\bar{z}_1 s_1 + \bar{z}_2
s_2)]f(t_1,t_2,s_1,s_2),\nonumber\\
&&(t_1+t_2)(s_1+s_2) = E
\label{5.3}
\eea
The idea is to integrate over one of  $t_1, s_1$ and one of
$t_2, s_2$ independently to satisfy proper monodromy, but the
constraint precludes mixed types involving $t$ and $s$, and it
will be sufficient to examine the $z$ type only. Since the third
variable turns out superfluous, we are led to the integrand
\be
e^S=\exp[z_1 t_1+z_2 t_2-(\bar{z}-\bar{b})E/(t_1+t_2)]%
t_1^{\al_11}t_2^{-\al_2-1},
\label{5.4}
\ee
with an arbitrary $\bar{b}$. The integration contours
$(C_1,C_2)$ are formally taken respectively around $0$ to
$U(-\theta_1))$ and $U(-\theta_2)$.  It is in general not
possible, however, to avoid the vanishing of $t_1+t_2$ by a proper
choice of the contours, and this would invalidate the formula. As
$C_1$ or $C_2$ makes a $2\pi$ rotation, there always occur
crossings of $-C_1$ and $C_2$,  and we have to set up a convention
for one of the $t$'s to make a detour around the other. A change of
the crossing point leads to a change in the integral, and after
making a $2\pi$ rotation, it will cause an extra change in $\psi$
beyond a phase factor. A detailed study shows that the extra piece is 
a regular function of $z$ and spoils the monodromy properties. 
Therefore it is not possible to satisfy the two monodromy conditions 
simultaneously. Later an alternative form of single integral 
representation for dynamical fluxes will be found to solve the problem. But 
we will first show next that the multiple integral representation can be 
used without difficulty if a magnetic field is present, athough the 
solutions obtained are of the type that do not have free field limits. 

\bigskip

\subsection{Solutions in a magnetic field}

Unlike the field-free case, it is easy to generalize the $z$-type 
representation for unshifted levels, Eq.(4.11), to $N$ fluxes with the multiple
 integral method. The solution is given by
\bea
\psi&=&\exp(-\ga z\bar{z}/2)\int \cdots \int_{C}\exp(\sum{z_i}
t_i)%
(T-\ga\bar{z})^n\prod(t^{-\al_i'-n-1}dt_i), \nonumber \\
T&=&\sum t_i,
\al_i' =\al_i+l_i>0,\hspace{1cm} i=1...N
\label{5.5}
\eea
The factor $(\sum(z_i t_i - \ga\bar{z})^n$ can become zero, but
this does not cause the earlier problem since it is a polynomial
in the $t_i$'s containing powers up to $N$. The condition on the
$\al_i$'s insures that the overall power of each $t_i$ is $<1$.
For the ground state $n=0$ we have
\be
\psi_{\{\al'\},0}=\prod {z_i}^{\al'}/(\al')!,\,\,
\Psi_0=\exp(-\ga z\bar{z}/2)\psi_0
\label{5.6}
\ee
The higher levels are generated by applying the raising operator
$\partial_z-\bar{z}\ga$ $n$ times to $\psi_{\{\al'+n\},0}$. Thus for
this class of solutions the Landau levels are not affected by the
fluxes. The wave function is a homogeneous polynomial of order $n$
made up of $\bar{z}\ga$ and the $z_i$'s.
 
\bigskip
 
\subsection{Zero field limit}

Earlier we found that the multiple integral representation
failed in the field-free case, so it is an interesting problem
to study the limit of letting $\ga$ go to zero,
$\ga =E/(n+1/2)\ra 0$, keeping energy fixed, or $n \ra \infty$.
We will evaluate the formula Eq.(5.5) for $N$ fluxes in the
$n \ra\infty$ limit by finding the saddle points with respect to
each $t_i$:
\be
t_i=(\al_i'+n)/(z_i+n/(T-E\bar{z}/n))\sim T, T=\sum t_i\sim NT
\label{5.7}
\ee
which is a contradiction for $N>1$. A more careful analysis
shows $t_i \sim O(n)$. The wave function is then $O(n^{(1-N)n})$, and
it is driven to zero. For $N=1$, on the other hand, a consistent
limit exists as was observed in section 4. By writing
\be
(\al_i'+n)/t_1 =(z_1+ n/t_1 + E\bar{z}/t_1^2),
t_1=\al_1'/(z_1+E\bar{z}/t_1^2)
\label{5.8}
\ee
we get the same equation derived from the zero field solution
Eqs.(2.8),(2.9). The origin of the problem is traced back to the
fact that each $t_i$ factor in Eq.(4.16) had a power ${-\al'-n}$ in order 
to give negative powers of $t_i$ for all the terms
$\sim T^m(\ga\bar{z})^{n-m}$ in $(T-\ga\bar{z})^n$, and this
contributed an overall power of $Nn$.
 
\bigskip
 
\sectionnew{\large Dynamical fluxes}
\subsection{Wave functions for the charge-flux system}
 
The difficulties of the general $N$-flux problem can be resolved if we treat
the fluxes themselves as dynamical objects with their own kinetic energies,
for reasons which are not immediately clear. First we note that, when the 
fluxes are made dynamical objects, they are under the influence of a magnetic 
(dual) vector potentials produced by the charges which, however, are of the 
same form as the electric vector potentials produced by the fluxes and acting 
on the charges. This is obvious because, for the wave function $\psi$ as a 
function of the coordinates of a charge and a flux, a rotation of the charge 
around a flux should be equivalent to a rotation of the flux around the charge.
(A general description of charges and fluxes as dual objects is given in the 
Appendix 2.) Thus the gauge transformation $G$ of Eq.(2.5) will remove the 
potentials from both the charge and the flux, and we will end up with a free 
Hamiltonian
\be
H=H_z+H_b=-\partial_z\partial_{\bar{z}}-(1/\mu)\partial_b\partial_{\bar{b}}
\label{6.1}
\ee
where the flux is represented by coordinates $b,\bar{b}$ and
mass $\mu$ (the mass of the charge $=1$). This generalizes to
any number of fluxes as well as charges if we consider only
charge-flux interactions, and ignore charge-charge and flux-flux
interactions.
 
The center of mass commutes with the free Hamiltonian, hence it
is a constant of motion even with the boundary condition imposed
by the flux singuralities since they depend on relative coordinates. So the  
wave function can be separated into that for a free center of mass motion 
and that for the relative motion, and we may deal with only the relative part. 
We will not do this here because it will complicate the formulation.

Suppose there are $N$ fluxes with masses $\mu_i,i=1..N$. We relabel the $2N+2$ 
coordinates of the charge and the dynamical fluxes 
$z, \bar{z}, \{b_i,\bar{b}_i\}$ uniformly as $\{u_i, \bar{u}_i\}, i=0..N$ and 
similarly the masses as $\{\mu_i\}, \mu_0=1$. The fluxes will eventually be 
set to their classical positions $b_{i}^0=u_{i}^0, i \neq 0$. The coordinates 
relative to them are denoted with primes: $u_i'=u_i-u_{i}^0, i>0)$. The total 
Hamiltonian $H$ is the sum of $N$ individual ones 
\be
H_i=-(1/\mu_i)\partial_{u_i}\partial_{\bar{u}_i}, \sum H_i = H
\label{6.2}
\ee
Further define
\be
I_i=\exp(-\mu_i u_i\bar{u}_i'/t)/t), I = \prod I_i
\label{6.3}
\ee
The $I$'s are the heat kernels satisfying the relations
\be
H_i I_i=(-\mu_i u_i\bar{u}_i/t^2+1/t)I_i=-\partial_t I_i,
H I = -\partial_t I,\,\,(t\neq 0)
\label{6.4}
\ee
Here $t$ may be regarded either as an imaginary time in a
time-dependent Scroedinger equation, or simply as a parameter to be
integrated over.
\footnote{With the first interpretation, there are time-dependent solutions for
non-dynamical fluxes of the following form, using only the kernel: 
$I_0(z):\psi=I_0\phi(z/t), \phi=\prod_{i=1\dot N}(z/t-iv_i)^{\al_i'}$. The 
flux coordinates $b_i=v_i t$ are moving with velocities $v_i$. But since the 
$\bar{b_i}$'s are fixed, it does not correspond to physically moving fluxes. 
(However, this form is of the same type as the function $\phi$ given below.) 
Solutions with dynamical fluxes of the simple form 
$\exp(iPX+i\bar{P}\bar{X})\prod(z-b_i)^{\al_i'}$, where $P$ and $X$ refer to 
the center of mass, are not admissible since they are not bounded at infinity.}

The ansatz for $\psi$ is now,
\be
\psi\equiv=\int_C \exp(Et)I\phi(u_i,t)dt,
\label{6.5}
\ee
with some contour $C$. Now apply $H$ to it, making use of Eq.(6.3):
\bea
H\psi&=&\int\exp(Et)[(-\partial_tI)\phi-\sum 1/\mu_i\partial_{\bar{u}_i'}I\partial_{u_i}\phi]dt\nonumber\\
&=&E\psi+\int(-\exp(Et)[\partial_t I)\phi+I\sum (u_i\partial_{u_i}\phi%
]dt\nonumber\\
&=&E\psi+\int\exp(Et)DI\phi dt, \nonumber\\
D&\equiv&t\partial_t+\sum u_i\partial_{u_i} 
\label{6.6}
\eea
So $\psi$ will be a solution provided that $D\phi=0$, i. e.\, $\phi(u_i, t)$ is
a homogeneous function of degree zero. We will choose
\footnote{There is a freedom of including an optional factor $z/t$ representing
a nondynamical flux at the origin. When no other fluxes are present, this is 
equivalent to Eq.(2.13). We also note that even a single flux, not located at 
the origin of coordinates, does not have a static representation.}
\be
\phi=\prod_{i=1\dots N}((u_0-u_i)/t)^{\al_i'}
\label{6.7}
\ee
with a contour $C=(-\infty,(0+))$.
Thus
\bea
\psi&=&\int\exp(Et-\sum u_i\bar{u}'_i/t)
\prod((z-b_i)/t)^{\al_i'}dt/t^{N+1} \nonumber\\
&=&J_{\sum\al'+N}(2RE^{1/2})R^{-N}\prod((z-b_i)/R) ^{\al_i'},\nonumber\\
R&=&(\sum\mu_i u_i\bar{u}_i')^{1/2}
\label{6.8}
\eea
up to a numerical factor. Its conjugate representation is obtained by 
replacing the $u_i, \al'$ by $\bar{u}_i, -\al_i'$, and changing the contour to
$(0, (0+))$ (or equivalently $t\ra 1/t$ with the original contour).
The above $\psi$ obviously has the right monodromy properties. When the flux 
masses become large, $\mu_i \ra \infty$, the Bessel function will become small 
and oscillate rapidly except at their intended `classical' positions where 
$E^{1/2}\mu_i >>|u_i'|\,\, (i \neq 0)$, $R^2\ra z(\bar{z})$. For large 
values of $|z|, \psi$ behaves like a single-flux solution with strength 
$\sum \al'+N$.  
\footnote{For $|z|>>|b_i|, \phi$ satisfies the truncated form of 
Eq.(6.6): $D'\phi\equiv(z\partial_z+t\partial_t)\phi=0$ at the saddle points, 
which assures, as may be seen easily, that $\psi$ is a semiclassical 
(Hamilton-Jacobi type) solution of $H_z$.}
In momentum space, $\psi$ has a simple but symbolic form
\be
\psi(k_i,\bar{k}_i)=\delta(\sum k_i\bar{k}_i/\mu_i-E)\prod (i(k_0-k_i))^{\al_i'}\label{6.9}
\ee

The Bessel function in Eq.(6.8) corresponds to a solution of the wave equation 
in a $(2N+2)$-dimensional space in the so-called hyperspherical harmonics 
expansion. In the absence of the factor $\phi$, the general solutions in this 
space are spanned by Bessel funcions times the $(2N+2)$-dimensional spherical 
harmonics. They can be generated by applying $\partial_{u_i}$ and 
$\partial_{\bar{u}_i}$, which commute with one another and with $H$, repeatedly
to the $s$-wave solution $J_N/R^N$. Obviously each operation respectively 
generates a negative or positive angular momentum around the origin in the 
$2$-dimensional subspaces spanned by $u_i, \bar{u}_i$. In the integral 
representation of Eq.(6.8) without $\phi$, application of $\partial_{u_i}$ 
($\partial_{\bar{u}_i}$) will bring down a factor $\bar{u}_i/t\,\, (u_i/t)$. 
Application of a monomial of these operators will generate a polynomial in the 
$u$'s and $\bar{u}$'s, and this way a complete set of states labeled by the 
$N+1$ `magnetic' quantum numbers of $O(2N+2)$ and a Casimir invariant (the 
degree of the monomial) will be generated. Fractional factors in  $\phi$ are of
a similar nature, but the power of angular momentum around each flux are 
correlated to be all positive for $\psi$, or all negative for its conjugate 
representation.  Nevertheless, by applying these same operators to the $\psi$ 
of Eq.(6.8), new admissible states can be created as long as the lowest power 
of each factor $(z-b_i)$ is positive. Together with their conjugates, these 
will presumably span a complete set of states.

The solutions of the type given by Eq.(6.8), however, are not the natural 
ones when considered as wave functions for each of the individual particles. 
Because of the factor $1/t^{N+1}$ characteristic of the 
$(2N+2)$-dimensional space, they go like 
$1/R^{N+1/2}\sim 1/|u_i|^{N+1/2}$ when one of the $u_i$'s becomes large, 
contrary to the expectation that they should go like $1/|u_i|^{1/2}$ in the 
physical $2D$ space. The extra power of $1/t$ which is responsible for 
this is a quantum effect associated with the dyamical degrees of freedom of 
each flux. If a flux is to be pinned to a fixed position, its degrees of 
freedom will have to be frozen. By the same token, if two fluxes are to be 
merged, those of relative motion will have to be suppressed. But the flux 
velocities $\partial_b/\mu, \partial_{\bar{b}}/\mu$ (unless evaluated at the 
origin) are independent of $\mu$, which was necessary for construction of the 
factor $\phi$. So the derivatives, i.e.\, the dynamics of the fluxes cannot be 
ignored at the level of the Schroedinger equation even for large masses. 
Herein lies the need for treating the fluxes as dynamical, and the 
difficulties of pinning them down. 
\footnote{However, in the opposite limit of the single charge becoming
infinitely heavy, the wave function of the fluxes would reduce to a product of 
the individual ones as in the case of many charges in the presence of a single 
static flux.}

A sensible way to recover the proper $2D$ asymptotic behavior may be to produce
highly polarized states in the direction of the particular pair of coordinates 
$u_i, \bar{u}_i$, i. e., those consisting of high powers of homogenious 
monomials $u_i/t$ and $\bar{u}_i/t$. But this cannot be a finite 
series since they increase powers of $1/t$ rather than reducing them.
Consider then applying the Green's function
\be
P=1/(E_z-H_z)=1/(\partial_z\partial_{\bar{z}}+E_z),\,\, E_z<E
\label{6.10}
\ee
to the $\psi$ of Eq.(6.8). $P$ commutes with $H$ so $P\psi$ is again a 
solution, and should go like $\cos(2E_z^{1/2}|z|+\delta)/|z|^{1/2}$. Since the 
source $\psi$ is concentrated near the origin ($\sim 1/|z|^{N+1/2}$), it is 
clear without explicit computation that  $P\psi$ will go as $1/|z|^{1/2}$. In 
order for this to be an exact solution for all $z$, however, $P$ must be a 
Green's function in which the flux singularities are already incorporated. 
Such a Green's function can in principle be constructed from a complete set of 
states, but it is not practical to carry it out. 

Instead, we can define an effective Hamiltonian for which solutions with the 
desired porperties can be written down easily:
\bea
H_{\rm eff}&=&H_z-z^{-1}\partial_{\bar{z}}\sum b_i\partial_{b_i}-\bar{z}^{-1}\partial_z\sum \bar{b}_i\partial_{\bar{b}_i}\nonumber\\
&=&-z^{-1}\partial_{\bar{z}}\sum_0 ^N u_i\partial_{u_i}-\bar{z}^{-1}\partial_z\sum_1 ^N \bar{u}_i\partial_{\bar{u}_i}\nonumber\\
&=&-\bar{z}^{-1}\partial_z\sum_0 ^N \bar{u}_i\partial_{\bar{u}_i}-z^{-1}\partial_{\bar{z}}\sum_1 ^N u_i\partial_{u_i}
\label{6.11}
\eea
with the stipulation that elementary solutions depend only on either the $b$'s 
or the $\bar{b}$'s, and satisfy proper monodromy and boundary conditions. 
Indeed, 
\bea
\psi&=&\int(-z\bar{z}/t+Et)\prod(z-b_i)/t)^{\al_i'}dt/t\nonumber\\
&=&J_{\sum \al_i'}(2E^{1/2}|z|)\prod(z-b_i)/|z|)^{\al_i'}
\label{6.12}
\eea
and its conjugate form clearly satisfy Eq.(6.11), and have the right 
aymptotic behavior. Being linear in the derivatives, we may say that
the flux coordinates retain only partially their dynamical status. The general 
properties of $H_{\rm eff}$ will not be exmanined here, but Eq.(6.12) will be 
adopted in subsequent sections.

\bigskip

\subsection{Solutions in a magnetic field}
 
The Hamiltonian for a charge in a magnetic field is
\be
H_z = -(\partial_z - \bar{z}\ga)\partial_{\bar{z}}
\label{6.13}
\ee
and a solution which has a zero-field limit is, according to
Eqs.(4.8),
\bea
\psi&=&\int \exp(zt)(1-\bar{z}\ga/t)^n dt/t \nonumber\\
&=&\int\exp(-z\bar{z}/t)(1+\ga t)^n dt/t =\int I_z dt/t\nonumber\\
I_z&=&\exp(-z\bar{z}/t)(1+\ga t)^n/t
\label{6.14}
\eea
The last expression was obtained by a change of variable
$t\ra -\bar{z}/t$. It satisfies the relation
\be
(H_z-n\ga)I_z = -\partial_t((1+\ga t)I_z)
\label{6.15}
\ee
By substituting the new $I_z$, the analog of Eqs.(6.4-5) becomes
\bea
\psi&=&\int I\phi dt,\nonumber\\
(H-E)\psi&=&\int-(\partial_t(1+\ga t)I)\phi
+\sum (u_i/t)\partial_{u_i}\phi,\nonumber\\
&=&\int I D'\phi dt/t, D'=(1+\ga t)t\partial_t+\sum u_i\partial_{u_i}
\label{6.16}
\eea
 
The solution for $\phi$ is then obtained by substituting
$t/(1+\ga t)$ for $t$ in Eq.(6.5), and $\psi$ becomes
(dropping $\exp(Et)$ in favor of $(1+\ga t)^n$)
\bea
\psi=\int\exp(-\sum u_i\bar{u}_i'/t)(1/t+\ga))^N(1+\ga t)^n\nonumber\\
\prod((z-b_i)(1/t+\ga))^{\al_i'} dt/t
\label{6.17}
\eea
The integrand has a branch point at $t=0$ of order $-N-\sum \al'$
and one at $t=-1/\ga$ of order $N+n+\sum \al'$. We can adopt a
closed contour $(-1/\ga+,0+)$ around them. When the integral is
evaluated by expanding the exponential, we get a series of Beta
functions
\be
(\ga r')^m B(-N-\sum\al'-m,N+n+\sum\al'+1)/m!
\label{6.18}
\ee
which is zero if $m \ge n+1$, so it is a polynomial of order $n$. In
the limit $\ga \ra 0, n=E/\ga\ra \infty$ for fixed $E$, $(1+\ga t)^n$
becomes $\exp(Et)$, the contour becomes $(-\infty, (0+))$, and
Eq.(6.17) reduces to that for the field-free case.
 
\bigskip
 
\subsection{Shifted Landau levels}
 
In section 4.2 we showed that there are shifted Landau levels for a
single static flux. Eq.(6.17) serves a basis for incorporating $N$
dynamical fluxes. Assume this time that $\ga<0$, replace
$n$ in Eq.(6.16) by $-n-\sum\al'-1$ so that the energy
$E/\ga = (n+\sum \al'+1/2)/|\ga|$ is still positive.
The singularity at $-1/\ga$ becomes a pole of order $N-n$. With a
change of variable $t=s-1/\ga$ the full wave function $\Psi$ can be
turned into
\bea
\Psi&=&\exp(-\ga z\bar{z}/2)\psi \nonumber\\
&=&\oint_{0+}\exp(-r'^2/(s-1/\ga)-\ga z\bar{z}/2)
\ga^{N-n-1}\prod((z-b_i)^{\al_{i}'}\nonumber\\
&&(s-1/\ga)^{-N-\sum\al'-1}s^{N-n-1} ds
\label{6.19}
\eea
which is nonzero if $n \ge N$, or if $n \ge 0$ after $N$ is
eliminated. $\Psi$ converges at $\infty$ since $\ga <0$ and
$r'=z\bar{z}$ effectively. The energy is $n+\al_i'$. This, then, is
the generalization of the shifted Landau levels, Eqs.(4.9), (4.10).
For a single static flux plus $N$ dynamical fluxes, the energy gets
shifted by all the flux strengths simultaneously. We do no know if partially
shifted states by some of the $\al_i'$s can be found. When 
$\ga \ra 0, n \ra \infty$, the integral is pushed to $\infty$ and disappears, 
meaning that the shifted states have no free field limits.
 
\bigskip
 
\sectionnew{\large Large N limit and flux medium}
 
According to the results of Section 6.1, we will effectively regard the fluxes 
as fixed objects and deal with wave functions of the type given in 
Eq.(6.12). The order of the Bessel function representing $\psi$ then goes up 
with the number of fluxes as $\sum \al_i'$, with all $\al$'s being of the same 
sign $>0$ and adding up. Suppose the fluxes of equal strength $\al$ are 
distributed with a density $1/a^2$ up to a large radius $R_0$. The total flux 
is $\al\pi(R/a)^2$. The angular momentum $L$ of the wave function at radius 
$R<R_0\,\,(R>R_0)$ is $\al\pi(R/a)^2 \,\,(\al\pi(R_0/a)^2$).
Its azimuthal momentum is $L/R = \al\pi R/a^2\,\,(\al\pi (R_0/a)^2/R)$, so the 
radial momentum for large $R$ :\,\, $\al\pi R^2 E>>1$, is 
$p_R \sim \pm i\al\pi R/a^2$, and the radial ``phase'' is 
$\int ip_R dr \sim \pm \al\pi(R/a)^2/2$ \,\,$(\pm (R_0/a)^2\ln R)$.
For $R<R_0$, the wave function will behave like Gaussian or anti-Gaussian. 
Below we will investigate it in more detail.
 
Assume that all fluxes have the same strength $\al$, and form a regular square 
lattice of unit size at sites $b_i, i=m+in$, although these assumptions are not
essential. The gauge potential $A_z,A_{\bar{z}}$ and the gauge function $G$ 
are
\bea
A_z&=&(\al/2)\sum 1/(z-b_i), \nonumber\\
A_{\bar{z}}&=&-(\al/2)\sum 1/(\bar{z}-\bar{b}_i),\nonumber\\%
G&=&\prod((\bar{z}-\bar{b}_i)/(z-b_i))^{\al/2}
\label{7.1}
\eea
As the sum goes to infinity, they fail to converge, hence an extra
pure gauge subtraction terms become necessary (except for $i=(0,0)$).
The modified forms are then
\bea
A_z&=&(\al/2)\zeta(z), A_{\bar{z}}=(-\al/2)\zeta(\bar{z}),\nonumber\\
\zeta(z)&=&\sum(1/(z-b_i)+1/z+z/b_i^2),\nonumber\\
G&=&[\sigma(z)/\sigma(\bar{z})]^{\al/2},\nonumber\\
\sigma(z)&=&z\prod_{i\neq 0}((1-z/b_i)\exp(z/b_i+z^2/2b_i^2))
\label{7.2}
\eea
Here $\zeta$ and $\sigma=\exp(\int \zeta dz)$ and are the elliptic
functions of Weierstrass with full periods $1,i$. Because of the subtractions, 
they are not periodic but satisfy
\bea
\zeta(z+m+in)&=&\zeta(z)+\pi(m-in),\nonumber\\
\sigma(z+m+in)&=&\sigma(z)\exp[\pi(m^2+n^2)/2+i\pi^2 mn/4+z\pi(m-in)]
\label{7.3}
\eea
The leading asymptotic factors are insensitive to the regular lattice
structure, and may be derived by a continuum approximation. If the
flux density is $1/a^2$,
\bea
\sum 1/(z-b_i)&\ra&\int 1/(z-r\exp(i\theta) r dr d\theta/a^2\nonumber\\
&=&\pi|z|^2/za^2=\pi\bar{z}/a^2
\label{7.4}
\eea
The contributions to the integral over $b$ from $r>|z|$ 
vanish because the phases of the contributions from points at radius $R>|z|$ 
from the origin cancel. Integrating this from $0$ to $z$ along the radius, we 
get
\be
z\prod'(z-b_i)/b_i)\equiv \sigma_R \sim \exp(\pi(|z|/a)^2/2)
\label{7.5}
\ee
These non-analytic expressions are a result of confusing $z$ and the cell 
coordinates $(m,n)a$. We can also use this method to define a cutoff version 
of $\zeta$ and $\sigma$:
\bea
\zeta_{R_0}(z/a)&\sim&\pi\bar{z}/a\hspace{5mm}{\rm if}\hspace{5mm}|z|<R_0,\nonumber\\
&=&0 \hspace{5mm}{\rm if}\hspace{5mm}|z|>R_0;\nonumber\\
\sigma_{R_0}(z/a)&\sim&\exp(N/2)\hspace{5mm}{\rm if}\hspace{5mm}|z|<R_0,\nonumber\\
\sigma_{R_0}(z/a)&\sim&\exp(N/2)\hspace{5mm}N=\pi R_0^2{\rm if}\hspace{5mm}|z|<R_0,\nonumber\\
&\sim&\exp(N/2)(z/R_0)^{N}\hspace{5mm}{\rm if}\hspace{5mm}|z|>R_0, 
\label{7.6}
\eea
We can now substitute them in Eq.(6.12), taking the origin of coordinates at 
the center of the lattice. 
\footnote{A possible alternative choice for $\phi$ is 
$\phi=\sigma_{R_0}(z/a)/\sigma_{R_0}(lt)$ with some $l$ and the contour running
outside of the fluxes: $lt>R_0$. But this leads to the same asymptotic results 
as those given below, up to a normalization.}
Then $\phi =\sigma_R/t^{\al N}$. First, the value of
$\psi$ at the origin can be exactly calculated, and yields $\psi(0)=1/N!$. 
For the asymptotic behavior, we can use the the saddle points to get
\be
\psi\sim C(t_0)\exp(-E|z|^2/t_0+t_0)t_0^{-N\al}\sigma_{R_0}(z/a)^{\al} 
\label{7.7}
\ee
where $C(t_0)$ is the contribution from the Gaussian integral around $t_0$, and
\bea
t_0&=&\pm iE^{1/2}|z| \hspace{5mm} {\rm if}\hspace{5mm} N>>E|z|^2,\nonumber\\
\psi&\sim&\exp(\pm 2iE^{1/2}|z|+N\al/2)(z/\pm iaE^{1/2}|z|)^{N\al}/(\pm iE^{1/2}|z|)^{1/2};\nonumber\\
t_0&=&N\al\hspace{5mm}{\rm if}\hspace{5mm}ER_0^2<E|z|^2<N,\nonumber\\
\psi&\sim&N^{1/2}\exp(3N\al/2)(z/aN\al)^{N\al} \nonumber\\
&=&\exp(N\al/2)(z/a)^{N\al}/N! \nonumber\\
&=&\exp(N\al/2)(z/a)^{N\al}\psi(0);\nonumber\\
t_0&=&N\al\hspace{5mm}{\rm if}\hspace{5mm}1<E|z|^2<ER_0^2,\nonumber\\
\psi&\sim&\exp(\al\pi |z|^2/2)(z/|z|)^{\al\pi|z|^2}\psi(0)
\label{7.10}
\eea
The first form is the true asymptotic limit. The second is valid in the 
near outside region $|z|>R_0$. The third applies to the inside of the flux 
medium. These are all consistent with the physical argument given at the
beginning of the section.
\footnote{In the exterior region, $\phi$ satisfies 
$D'\phi=z\partial_z\phi=-t\partial_t\phi=\al N$, so $\psi$ is a semiclassical 
solution of $H_z$ (footnote 5), whereas $D'\phi$ is not zero in the interior 
region. (Of course the full $D\phi=0$ is always guaranteed).}
From above we see that the wave function inside the flux area grows with the 
distance in an anti-Gaussian way like $\exp(\al\pi|z|^2)$ inside the medium, 
then with a power like $|z|^{N\al}$ to distances $E^{1/2}|z| > N\al$ where the 
true asymptotic behavior sets in. In other words, the wave function is pushed 
out ot the flux region and beyond (unless the energy is so large as to skip the
intermediate regimes).

\bigskip
 
\sectionnew{\large Summary and discussion}
 
In terms of physical concepts, the Aharonov-Bohm problem for many
fluxes is a straightforward extention of the single flux case, and
is not expected to contain anything conceptually new. Yet in an
attempt to construct explicit solutions, we have found some
unexpected features. In the idealized A-B problem of an infinitely
thin flux, the wave function of a charge in its vicinity is
described by Bessel functions whose order is shifted by the flux
strength $n$ to $|\al+n|$. Since the latter has to vanish at
the flux, the component which would be an $s$-wave in free space is
pushed out irrespective of the sign of $\al$. This effect will
magnify in a medium made up of fluxes.
  
To handle the general case of many fluxes, the gauge potential may
be removed by a singular gauge transformation. The Hamiltonian is
reduced to that of a free particle except for boundary conditions.
Around each flux $i$ the transformed wave function must develop a branch 
point in one of the complex coordinates, with a positive fractional power 
so that it has the form $z^{\al_i+n_i}$, or $\bar{z}^{-\al_i+n_i}$, times 
a holonomic function in $z$ and $\bar{z}$. In particular, at zero energy the 
general solution is an arbitrary sum of a function of $z$ and a function of 
$\bar{z}$, and each part must carry all the branch points. In other words, the 
fractional angular momentum around each flux, as opposed to the intrinsic flux 
strengths $\al_i$, must all be either positive for the analytic, and negative 
for the anti-analytic solution. At nonzero energies, the wave function of 
eithertype contains both variables, but by continuity the above qualitative 
features do not change. (This turns out to be true even in a magnetic field 
where the spectrum is discrete.) The total angular momentum of a wave function 
around a large circle containing a cluster of fluxes is then a sum of positive 
or negative numbers which will grow with the number of fluxes, resulting in the
expulsion of a charge from the interior of the cluster. This is because the 
wave function has to vanish at each flux site, whether the fluxes add up to an 
integer or not. Only if an individul flux is integral, will it not contribute 
to the effect. This conclusion does not require an explicit construction of 
solutions. 
 
These general results do not seem to be due to our reliance on the singular 
gauge and analyticity, or the simplifying assumption of infinitely thin fluxes. To see this, first consider a flux $\al_1$ at the origin in the original gauge 
so a solution may be expanded in Bessel functions $J_{|\al_1+n|}$ of positive 
fractional order. If there is a second flux $\al_2$ at point $P$, they have to 
be re-expanded in Bessel functions $J_{|\al_2+m|}$ around $P$. But the complete
set of eigenfunctions for the latter covers only the space $R^2-P$. Therefore 
we have to start from the beginning with a function that has the proper 
behavior at both sites. (The general addition formulas for Bessel functions 
show that all positive and negative powers and with the same $\al$ are 
involved in the re-expansion around $P$ if $\al$ is not an integer. The 
integer case is special since $J_{-n}=(-1)^n J_n$.)
 
It is also instructive to compare the A-B problem with that of hard core
potentials. Here the similarity is that the eigenfunctions do not form
a complete set in $R^2$. The difference is that in the latter the
eigenfunctions are Bessel functions of integer order, the $s$-wave included. 
If the radius of the potential is finite, the solutions involve both regular 
functions $J_n$ and singular ones $N_n$. This is equivalent to having both 
positive and negative orders for noninteger $n$, or the
Hankel functions $H_{\al+n}^{1,2}$. In a realistic A-B problem where
the fluxes have a finite size\cite{Lewis}\cite{Sakoda}, the solution around a 
flux will be a superposition of functions $J_{\al+n}$ with all $n$, or, after 
gauging away the potential, it will have all powers
$z^{\al+n}$ or $\bar{z}^{-\al-n}$. This, however, would not alter the 
difficulties with many fluxes since the quasi-free wave function has to satisfy
not only the condition that it vanishes at the boundaries but also that it has
the right fractional angular momenta in the neighborhood of the fluxes, in 
contrast to the hard core case where only the boundary condition matters.

The explicit construction of wave functions for many fluxes has also turned 
out to be possible, with or without a magnetic field. To construct a solution 
in free space, however, the fluxes have to be treated as dynamical objects 
with their own Hamiltonians. Technically we may say that it is because
kinematic correlations are induced between fluxes indirecty through those 
between the charge and the fluxes, and are independent of the masses. But the 
role of the  magnetic field in this context is unclear. These are puzzling 
features that remain to be understood in simple physical terms. Explicit 
solutions with proper asymptotic behavior can be witten down for an effective 
Hamiltonian which partially retains the dynamical nature of the fluxes. With a 
magnetic field, it is noteworthy that some of the Landau states get the energy 
shifted by the fluxes in proportion to the flux strength while others remain 
unaffected. The effect is asymmetric for the lowest level with respect to the 
relative sign of the field and the flux. This nonlinear behavior may lead to 
some interesting consequences. 

Acknowledgement. This work was supported by NSF Grant 9600697
and the University of Chicago. I thank Prof. T. Eguchi for
continuing interest and critical discussions (which started in 1979), and 
H. Awata, Y. Hosotani, R. Jackiw, H. Kawai, Y. Ohnuki, M. Peshkin, P. Ramond, 
P. Wiegmann and F. Wilczek for valuable comments and information about the 
literature at various stages of the work. Part of the work was done at Osaka 
University. I thank Prof. K. Kikkawa for his hospitality.
 
\bigskip
 
\renewcommand{\theequation}{A.\arabic{equation}}
\setcounter{equation}{0}
\section*{\large Appendices}
 
\bigskip
 
\subsection*{A.1 Complex metric}
 
Let a covariant and a contravariant vector be denoted in general
by $v_{.}$ and $v^{.}$ respectively, and let the contravariant
vector $v^{.} = (x,y)$ be mapped to $(z,\bar{z}) = (x+iy, x-iy)$.
The metric for the complex pair is then off-diagonal:
\be
g_{ik} = (1/2)[[0,1],[1,0]],\,\, g^{ik} = 2[[0,1],[1,0]],
\label{A1}
\ee
 
The rule for forming a vector $A$ in the complex metric is
\bea
&&A^{.} = (A^z, A^{\bar{z}})
= (A^x + iA^y, A^x - iA^y), \nonumber \\
&&A_{.} = (A_z, A_{\bar{z}} = (A^x - iA^y, A^x + iA^y)/2
\label{A2}
\eea
 
Since the gradient operator is
\be
\grad = (\partial/\partial{z}, \partial/\partial{\bar{z}}),
\label{A3}
\ee
and the Laplace operator becomes
\be
\nabla^2 = 2\{\partial/\partial{z}, \partial/\partial{\bar{z}}\}
\label{A4}
\ee
 
Rotation $R$ by an angle $\theta$ is  represented by a diagonal
matrix
\be
R_{.} = (\eta^{-1},\eta),  \, R^{.} = (\eta, \eta^{-1}) , \, \eta =
\exp(i\theta)
\label{A5}
\ee
so the outer product of two covariant or two contravariant vectors
is given by
\bea
&&v_{.} \times u_{.} = i(v_z u_z, -v_{\bar{z}}u_{\bar{z}}),
\nonumber \\
&&v^{.} \times u^{.} = -i(v^z u^z, -v^{\bar{z}}u^{\bar{z}})
\label{A6}
\eea
 
It follows also that the divergence and curl ($= F_{z\bar{z}}$) of
$A$ are given respectively by
\bea
&&{\rm div}\,A=(\partial_z A_z + \partial_{\bar{z}} A_{\bar{z}})/2,
\nonumber \\
&&{\rm curl}\,A_{.} = -{\rm curl\,}A^{.} = i(\partial_z A _{\bar{z}} -
\partial_{\bar{z}} A_{z})/2
\label{A7}
\eea
 
The Gauss' and Stokes' formulas read (on shell)
\bea
&&\int{v_n ds} = - i\int(v_{\bar{z}}dz - v_z d\bar{z}),
\nonumber\\
&&\int{v_s ds} = \int(v_z dz + v_{\bar{z}} d\bar{z})
\label{A8}
\eea
respectively.
 
The Laplace operator applied to the Green's function
\be
G = \ln{|z|}= (\ln z + \ln \bar{z})/2
\label{A9}
\ee
is zero, but the delta function singularity at the origin of the
on-shell space can be seen  from the Gauss' formula
\be
\int{\grad\,G_n ds} = (-i/2)\int(dz/z - d\bar{z}/\bar{z})= 2\pi
\label{A10}
\ee
\bigskip
 
\subsection*{A.2 Field theory description of charges and fluxes}
 
The fluxes can be regarded as dynamic magnetic particles. They
interact with the charges through the Aharonov-Bohm effect, but
not directly among themselves. A flux $\Phi$ is a concentration
$\Phi= Ba^2$ of the magnetic field density $B$ over an
infinitesimal area $a^2$, its energy (per unit $z$ direction) is
a part of the energy density $F_{xy}^2/2$ of the magnetic field plus
an amount necessary to keep it concentrated.\cite{Peshkin2} We will
assume uniformity and ignore the dynamics in the $z$ direction. When
a flux moves, the magnetic field inside will acquire electric field
components of order $v/c$, so $(B_z,-E_y,E_x)$ will be associated
with a moving flux. The vector potential associated with $E$ will be
present in the medium, even though it may be a shielded potential.
 
Let $A$ be the usual vector potential, $G$ and $F$ two vector
fields, $J$ and $K$ the electric and magnetic currents in $2+1$
dimensions, so they have mass dimension 2. $J$ is then not the usual
charge density since the charge is not extended in the $z$
direction, but may be thought of as a kink soliton
$\partial_z \sigma$ of an extended field $\sigma$. This is the
only place where the $z$ direction enters. Let the Lagrangian $L$ be
\bea
L&=&-\sum G_i\cdot F_i/2 - iG\cdot{\rm curl}A + eA \cdot J +gG \cdot
K+L_{mat},\nonumber\\
G_1&=&D_2,\,\, G_2= -D_1,\,\,G_3 = H_3,\,\,F=\Lambda G, \nonumber\\
\Lambda&=&{\rm diagonal\,matrix}\,[\lambda, \lambda, \kappa]
\label{A11}
\eea
$L_{mat}$ refers to the kinetic part of `matter' (charge and flux)
Lagrangian. Euclidean metric is used for simplicity, with $x_3=ix_0$,
etc. The electric and magnetic charges $e$ and $g$ are dimensionless;
$G$ and $F$ have mass dimensions $1$ and $2$ respectively. $B,H,E,D$
have the usual meaning in the Maxwell theory except for
dimensionality: the parameters $\kappa, \lambda$ are
respectively the analog of magnetic permeability and inverse
dielectric constant with the dimensions of mass. Eq.(A11) shows that
the magnetic ``vector potential '' is invariant under the usual
gauge transformation of the electric vector potential $A$. Varying
$G$ and $A$ as independent fields, we get the field equations
\be
F+i\,{\rm curl}\,A = gK, \hskip10mm i\,{\rm curl}\,G = eJ,
\label{A12}
\ee
If the magnetic current $K$ is absent, Eq.(A12) takes the usual
Maxwell form. If, on the other hand, $\kappa$ and $\lambda$ are
set to zero, the equations show a symmetry under duality between
$J,A$ and $K,G$:
\be
i\,{\rm curl}\,A = gK,  \hskip10mm i\,{\rm curl}\,G = eJ,
\label{A13}
\ee
and an invariance under both an electric and a magnetic gauge
transformation:
\be
A \ra A + {\rm grad}\,\phi,\,\, G \ra G + {\rm grad}\,\chi
\label{A14}
\ee
(Boundary contributions are assumed to vanish uder these gauge
transformations.) The left-hand side of Eq.(A13) are the
Coulomb-Lorentz forces acting respectively on charge and flux, which
are seen to come directly from the currents of their opposite
numbers.
 
For static electric and magnetic sources located respectively at
$x_j, y_j$ and $x_k, y_k$, Eq.(A13) yields
\bea
&A_1, A_2=g(-v_2,v_1)/(\pi|v|^2),v=(x-x_k, y-y_k)\nonumber \\
&G_1, G_2=e(-u_2,u_1)/(\pi|u|^2), u=(x-x_j, y-y_j) \nonumber \\
&A_{+}=-iA_1+A_2=g/\bar{z}, A^{-}=iA_1+A_2=g/z, z=v_1+iv_2, \nonumber \\
&G_{+}=iG_1-G_2=e/\bar{w}, G_{-}=iG_1-G_2=-e/w, w=u_1+iu_2
\label{A15}
\eea
Hence the vector fields $eA$ and $gG$ acting on charge and on
flux are the same, and proportional to $eg$, and the same form of
gauge transformation removes the potential from the charge and the
flux Hamiltonian.
 
If $\Lambda \neq 0$, there is no duality symmetry.  Conservation of
the magnetic charge requires
\be
{\rm div}\, \Lambda\,G = 0
\label{A16}
\ee
which is the same as in the Maxwell case. The magnetic gauge invariance
still holds if the $\chi$ in Eq.(A14) satisfies
\be
{\rm div}\,\Lambda\,{\rm grad}\, \chi = 0
\label{A17}
\ee
By taking the Coulomb gauge
$\partial_1 A_1 +\partial_2 A_2=0$, Eq.(A13) leads to
\bea
&\nabla^2 A_3=-ig\lambda^{-1}(\partial \times K)_{12}-\lambda
eJ_3,\nonumber\\
&\lambda^{-1}\nabla^2 A_1+\kappa^{-1}\partial_3 F_{31}=-
ig(\lambda^{-1}\nabla_2 K_3+\kappa^{-1}\partial_3 K_2)-
ieJ_1,\nonumber\\
&\lambda^{-1}\nabla^2 A_2-\kappa^{-1}\partial_3 F_{32}=-
ig(\lambda^{-1}\nabla_1 K_3-\kappa^{-1}\partial_3 K_1)-ieJ_2
\label{A18}
\eea
If static, $\partial_3 = 0$, or if $\kappa >>\lambda$, the
contribution to $A$ from $K$, and one to $G$ from $J$ become
\bea
\nabla^2 A_3&=&-\lambda eJ_3-ig\lambda^{-1}(\partial \times K)_{12},
\nonumber\\
\nabla^2 A_1&=&-ig\partial_2 K_3, \nabla^2 A_2 = ig\partial_1
K_3,\nonumber\\
G_3&=&i\kappa^{-1}g(\partial\times A)_{12}, \nonumber\\
G_1&=&i\lambda^{-1}\partial_2 A_3,G_2=-i\lambda^{-1}\partial_1 A_3
\label{A19}
\eea
Since the Green's kernel is $\ln(r)/2\pi, r=\surd(x^2+y^2)$, the
potentials $A_{1,2}$ and $G_{1,2}$ are then found to be the same as
Eq.(A13):
\be
eA_3=ieg\ln(r)/\pi,\,\, eA_{1,2}=-gG_{1,2}=
eg(\partial_2,-\partial_1)\ln(r)/\pi
\label{A20}
\ee
 
Eq.(A18) shows that the magnetic `Coulomb' potential $G_3$
acting on magnetic charge $gK_0=-igK_3$ is, apart from self
interaction, indeed the usual magnetic field
$G_{12}=H_3=B_3/\kappa$, where $\kappa$ is interpreted as an
effective magnetic permeability. The
term $G_{12} F_{12}/2=\kappa H_3^2/2$ is the magnetic field
energy. When an external magnetic field $H_{ex}$ is imposed, we
should change it to $\kappa(H+H_{ex})^2/2$, which means that the
flux will feel a magnetic Coulomb potential $G_{ex}$. By the
same token, if an external electric field is present without
suffering shielding, the flux will feel an electric Lorentz
force. On the other hand, the potential $A_3$ acting on a charge
at $(x,y)$ and generated by another charge at $0$ is
$-\lambda(\ln(r)/2\pi)eJ_3$. Since this should be the 3-D Coulomb
potential $1/(4\pi r)$, we must interpret $\lambda$ to be
$\sim \partial_z$, i.e.\, the derivative in the hidden $z$
direction at $z=0$.
\footnote{As was mentioned above, charge is like a kink soliton, or
a monopole with a string, that is sitting in the 2-D space. If
charges were extended in the $z$ direction like the fluxes, the
logarithmic potential would arise by integrating the 3-D Coulomb
potential over $z$:$\int dz/(r^2+z^2)= \ln((z+(r^2+z^2)^{1/2}/r)$
(after renormalizing away infinities). We get back the 3-D
potential if we take this as the 2-D potential and differentiate
it by $z$ at $z=0$.}

\end{document}